\documentstyle[manuscript,aps]{revtex}

\newcommand{\lam}{\lambda}
\newcommand{\eps}{\varepsilon}

\newcommand{\beq}{\begin{equation}}
\newcommand{\eeq}{\end{equation}}
\newcommand{\ba}{\begin{array}}
\newcommand{\ea}{\end{array}}
\newcommand{\beqa}{\begin{eqnarray}}
\newcommand{\eeqa}{\end{eqnarray}}
\newcommand{\bd}[1]{ \mbox{\boldmath $#1$}  }
\begin{document}

\draft
\title{Molecular Collective Vibrations in the Ternary Neutronless
Fission of $^{252}$Cf}
\author{\c S. Mi\c sicu$^{1,2}$\footnote{misicu@theor1.theory.nipne.ro}, 
P.O.Hess$^{3}$, A. S\u andulescu$^{1,2}$ and W. Greiner$^2$}
\address{$^1$ National Institute for Nuclear Physics, Bucharest-Magurele, 
P.O.Box MG-6, Romania}
\address{ $^2$ Institut f\" ur Theoretische Physik der J.W.Goethe
 Universit\" at, Frankfurt am Main, Germany}
\address{$^3$ Instituto de Ciencias Nuclerares, UNAM, Circuito Exterior, 
C.U.,  A.P. 70-543, 04510 M\'exico, D.F., Mexico}
\maketitle

\begin{abstract}

Based on a recent experimental finding which may suggest the existence
of a tri-nuclear molecular structure before the 
cold ternary fragmentation of $^{252}$Cf takes place, we solved the 
eigenvalue problem  of a certain class of vibrations which are 
very likely to occur in these molecules. 
These oscillations are the result of the joined action of 
rotations of the heavier fragments and the transversal vibrations of the 
lighter spherical cluster with respect to the fission axis. 
In the calculation of the interaction between the heavier fragments we took 
into account higher multipole deformations, including the hexadecupole one,
and introduced a repulsive nuclear part to insure the creation of a potential 
pocket in which a few molecular states can be accommodated.  The possibility
to observe the de-excitation of such states is discussed in connection with 
the molecular life-time. 
 
PACS number : 21.60.Gx,24.75.+i,25.85.Ca;
\end{abstract}

keywords : giant molecule, cold fission, collective motion, heavy-ion 
interaction

The scope of this letter is to extend our recent investigations on
the molecular configurations in the binary cold fission 
\cite{msg97,misi99} and the ternary cold fission \cite{hess99}. 
Growing interest aroused in the last year due to the experimental
indication of a long-living ($\geq 10^{-13}$s) structure in the $^{10}$Be 
accompanied ternary cold fission \cite{prl98}. Very recently a molecular 
structure in which $^{12}$C plays the role of the light accompanying
particle has been reported \cite{hwang99}. 

In this letter we make the suggestion that the ternary  cold fission of 
$^{252}$Cf is a process consisting of two main stages : in the 
preformation stage a quasi-bound molecular structure is formed in which
the heavier fragments are almost co-linear and the light particle 
(e.g. $\alpha$, $^3$Li ,$^{10}$Be and $^{12}$C) which is responsible for the 
molecular bonding, is orbiting in the equatorial region. 
This is similar to the case encountered in Molecular Physics, where in a 
linear or nonlinear chain of three atoms, the central atom ensures two 
bondings with the eccentric atoms \cite{wdc55}. In the second stage the 
quasi-molecular state is decaying \cite{smcg99}. Our aim is to study the 
collective vibrations of the system before the decay takes place.

Recently it has been advocated by us \cite{msg97}, based on the
concept of {\em nuclear molecule}\cite{hg84}, that for fragments
emitted in the binary cold fission, with almost no excitation energy, 
a collective vibrational spectrum will show up as a consequence of 
small non-axial fluctuations at scission.  
Such a molecular spectrum can be achieved if the interplay between the Coulomb 
and the repulsive nuclear core on one hand and the attractive nuclear part
on the other hand will produce a pocket in the interaction potential between 
the fragments\cite{seiw85}.  In the case of di-nuclear systems it was 
shown that possible molecular collective modes can be associated to the 
elongation variable and rotational vibrations taking place perpendicularly to 
the fission axis \cite{ns65}. The last type of modes, e.g. butterfly (bending) 
and anti-butterfly (wriggling) is also believed to be 
responsible for the formation of angular momenta in fragments emerging 
in binary spontaneous fission \cite{misi99,jor69,ziel74}.

In a previous paper the classical expression of the tri-nuclear Hamiltonian 
has been worked out for the case of the $^{96}$Sr+$^{10}$Be+$^{146}$Ba 
molecule in terms of the Jacobi variables $\bd{R}, \bd{\xi}$ and 
the angular velocities $\bd{\omega}'$ of the molecular frame 
\cite{hess99}. 
The equilibrium configuration was that of three aligned clusters, with the 
lighter in-between. In such a configuration the interaction between the 
heavy fragments is almost entirely given by the Coulomb term. However the 
interaction between the lighter fragment and the heavy fragments consists also
of a noticeable nuclear component, which in fact is responsible for the 
nuclear bond. Like in the case of binary molecules, butterfly modes can occur, 
in which the fragments rotates in phase while the lighter fragment is
approximately preserving its pole-pole configuration with the heavier 
fragments.   

The classical expression of the kinetic energy of the three-body
system, after removing the center of mass contribution, is expressed
as a sum of translational 
and rotational degrees of freedom :
\beq
T = {1\over 2}\mu_{12}{\dot{\bd{R}^2}} 
+ {1\over 2}\mu_{(12)3}{\dot{\bd{\xi}^2}}
+ {1\over 2}{^t}\bd{\omega}_1\bd{\cal J}_1\bd{\omega}_1
+ {1\over 2}{^t}\bd{\omega}_2\bd{\cal J}_2\bd{\omega}_2
+ {1\over 2}{^t}\bd{\omega}_3\bd{\cal J}_3\bd{\omega}_3
\label{kinetic}
\eeq
The first term describes the relative motion of the di-nuclear sub-system
(12) with reduced mass $\mu_{12}=m_1m_2/(m_1+m_2)$, whereas the second 
one corresponds to the relative motion of the third cluster with respect 
to the heavier fragments center-of-mass with reduced mass 
$\mu_{(12)3}=(m_1+m_2)m_3/(m_1+m_2+m_3)$.
The vectors $\bd{\omega}_{1,2,3}$ denote the angular velocities of the
rotational motion of the three clusters, referred to the laboratory frame, 
$^t\bd{\omega}$ being the transpose of  $\bd{\omega}$. 
In this paper we consider a spherical light cluster and thence the last 
term in eq.(\ref{kinetic}) disappears. The inertia tensors 
$\bd{\cal J}_i$ 
are defined in the intrinsic frame such that the only non-vanishing components 
are the first two diagonal terms, $(\bd{\cal J}_i)_{11}=(\bd{\cal J}_i)_{22}
\equiv J_i$,
the quantum rotation around the symmetry axis of any of the two heavier
fragments being discarded.

In what follows we are interested in studying the collective spectrum 
which develops upon constraining the tri-nuclear molecule to perform 
an oscilllatory motion similar to the {\em valence angle 
bending} in molecular physics and the butterfly(bending) modes in di-nuclear 
molecules, i.e. to perform small displacements from the equilibrium position 
which result in the decrease of the angle between the two valence bonds, 
$\Phi=\pi-\varphi_1-\varphi_2$, attached to the spherical light fragment 3
($\varphi_i$ is the angle between the axis joining the two heavier fragments 
and the line joining the heavy fragment $i$ with the light cluster). 
In the same time, since the nuclear proximity forces have the tendency
to keep constant the reciprocal distances and orientations of the
heavy fragments with the light one, we exclude possible 
{\em bond stretching} vibrations.  
If the bond stretching is absent, then there will be a corresponding 
decrease in the distance between the heavy nuclei 1 and 2, when the bending 
angles $\varphi_1$ and $\varphi_2$ are increasing.    
The quantitative translation of the above mentioned considerations  provides
us with a set of constraints between the variables of interest in this 
problem : $R$-the distance between the centers of the two heavier fragments, 
$\xi$-the distance between the light cluster 3 and the center-of-mass
of the heavy fragments ensemble, and the small bending angles $\varphi_1$, 
$\varphi_2$.
These last two variables are related between them, due to the assumption on 
the constancy of the pole-pole configuration between the light cluster 3 and  
the heavy fragments 1 and 2
\beq
\varphi_2 = \frac{R_1+R_3}{R_2+R_3}\eps
\label{defeps}
\eeq 
where $\eps=\varphi_1$.
Consequently we obtain the following relations, which allows us to eliminate 
from the kinetic energy (\ref{kinetic}) the variables $R$ and $\xi$ in
favor of $\eps$
\beqa
R & = & (R_1+R_2+2R_3)\left ( 1-{1\over 2}\frac{R_1+R_3}{R_2+R_3}\eps^2\right )\\
\xi & = & \xi_0 + 
{1\over 2}\left ( \frac{\partial^2 \xi}{\partial\eps^2} \right )_{\eps=0}\eps^2
\eeqa
where
\beqa
\xi_0&\equiv & \xi(0) = \frac{A_1(R_1+R_3)-A_2(R_2+R_3)}{A_1+A_2}\\ 
\left ( \frac{\partial^2 \xi}{\partial\eps^2} \right )_{\eps=0}
& = &
{A_1A_2\over A_1+A_2}
\frac{(R_1+R_3)(R_1+R_2+2R_3)^2}
{(R_2+R_3)(A_1(R_1+R_3)-A_2(R_2+R_3))}
\eeqa
Note that the above expressions have been written in the second order in
$\eps$. 

With the above choice the heavy fragments are constrained to rotate only 
around an axis perpendicularly to the axis joining their centers. 
This possibility is justified experimentally by the small forward anisotropy 
of the angular distribution of prompt $\gamma$ radiation.

As well as the above approximations we consider that the nuclei, 
building-up the molecule, are not performing $\beta$ or $\gamma$ vibrations.

We define a molecular frame whose $z$-axis coincides with the
fission axis and the three-body plane is choosen to coincide with the
$x-z$ molecular plane (in this way we eliminate the $y$-component of the
Jacobi coordinate $\xi$). Also, some assumptions have been made for the 
Euler angles of the interacting deformed fragments ($\chi_i,\varphi_i,\phi_i$).
The angle $\varphi_{i}$ has already been defined above : it describes
the angle between the fragment $i$ symmetry axis and the
molecular $z$-axis and it is expressed in terms of $\eps$ 
(see eq.(\ref{defeps})). The geometry of our problem, with the heavier 
fragments symmetry axes lying in the same plane, makes the Euler
angles  $\chi$ equal , i.e. $\chi_1=\chi_2$. They are combined in the
variable $\theta_3=(\chi_1+\chi_2)/2$ which measures the rotation
of the tri-nuclear aggregate with respect to the fission axis.

Following the standard procedures \cite{hg84,eisgre87} the total kinetic 
energy (\ref{kinetic}) can be expressed as a sum of three parts, the 
rotational energy $T_{rot}$, the internal 
kinetic energy $T_{int}$ and the Coriolis coupling $T_{cor}$.   
In terms of the time derivatives of the Euler angles,  
($\dot\theta_1,\dot\theta_2,\dot\theta_3$), specifying the 
rotation of the molecular frame,  the classical,  rotational 
kinetic energy reads: 
\beq
T_{rot}={1\over 2}\sum_{ij}g_{ij}^{rot}\dot\theta_i\dot\theta_j
\eeq
where the only non-vanishing components of the rotational metric
tensor $g_{ij}^{rot}$ are given by :
\beqa
g_{11}^{rot}&=&(J_0+J_1+J_2)\sin^2\theta_2 -  
J_{13}\sin2\theta_2\cos\theta_3+
\left (J_1+J_2\frac{R_1+R_3}{R_2+R_3}\right
)\eps\sin2\theta_2\cos2\theta_3
\nonumber\\
g_{22}^{rot}&=&J_0+J_1+J_2 \nonumber \\
g_{12}^{rot}&=&\left \{
J_{13}-2\eps\cos\theta_3\left(J_1+J_2\frac{R_1+R_3}{R_2+R_3}\right )
\right \}\sin\theta_3\cos\theta_2\nonumber\\
g_{23}^{rot}&=&\left\{ J_{13}-
2\eps\cos\theta_3\left(J_1+J_2\frac{R_1+R_3}{R_2+R_3}\right 
)\right\}\sin\theta_3\nonumber\\ 
g_{13}^{rot}&=&-\left\{
J_{13}\cos\theta_3-\eps\cos2\theta_3\left(J_1+J_2\frac{R_1+R_3}{R_2+R_3}\right
)\right\}\sin\theta_2
\eeqa
where $J_0=\mu_{12}(R_1+R_2+2R_3)^2+\mu_{(12)3}\xi_0^2$ and 
$J_{13}=\mu_{(12)3}(R_1+R_3)\xi_0$
  
The intrinsic kinetic energy will be comprised of ''$\eps$''-vibrations 
and intrinsic rotations of the clusters:
\beqa
T_{int}& = &{1\over 2}\left (
J_{13}\frac{R_1+R_3}{\xi_0}+J_1+J_2\left (\frac{R_1+R_3}{R_2+R_3}\right )^2\right )
\dot\eps^2
-J_{13}(\sin\theta_2\sin\theta_3\dot\theta_1+\cos\theta_3\dot\theta_2)\dot\eps
\nonumber\\
 &+ & {1\over 2}\left ( J_1+J_2\frac{R_1+R_3}{R_2+R_3}\right 
)\eps^2\dot\theta_3^2 \eeqa
There will be also a contribution from the Coriolis term
\beq
T_{cor}=\left(J_1+J_2\frac{R_1+R_3}{R_2+R_3}\right )
\left\{\eps(\sin\theta_2\cos2\theta_3\dot\theta_1-\sin 2\theta_3\dot\theta_2)
\dot\theta_3+(\sin\theta_2\sin2\theta_3\dot\theta_1+\cos 2\theta_3\dot\theta_2)
\dot\eps\right\}
\eeq

The next development of our considerations is facilitated by the 
peculiarities of the spontaneous fission of $^{252}$Cf. Since the spin
$\bd{J}$ of the mother nucleus is 0, the total helicity $K$, i.e. 
the projection of the total angular momentum on the fission axis is
zero for both binary and ternary fragmentations 
and therefore 
$\dot\theta_3=0$. To simplify even further one choose the molecular
frame such that $\theta_3=0$. 
After quantizing the kinetic energy in three coordinates 
$(\eps,\theta_1,\theta_2)$ and neglecting terms multiplied by the 
non-diagonal matrix-element  $J_1+J_2\frac{R_1+R_3}{R_2+R_3}-J_{13}$,
which prove to be small in the resulting metric tensor, we arrive to a form of 
the kinetic energy in which the rotations are decoupled from the 
butterfly vibrations
\beq
{\hat T}=-\frac{\hbar^2}{2(J_0+J_1+J_2)}\left ( \frac{1}{\sin\theta_2}
\frac{\partial^2}{\partial\theta_1^2}
+\cot\theta_2\frac{\partial}{\partial\theta_1}+
\frac{\partial^2}{\partial\theta_2^2}\right)
-\frac{\hbar^2}{2{\cal J}_{\eps}}\frac{\partial^2}{\partial\eps^2}
\eeq  
where 
\beq
{\cal J}_{\eps}=J_1+J_2\left (\frac{R_1+R_3}{R_2+R_3}\right )^2+
J_{13}\frac{R_1+R_3}{\xi_0}
\eeq

Next we turn our attention to the computation of the potential.
The total interaction energy is given by the sum 
\beq
V = \sum_{i\neq j=1}^3 V_{ij}(\bd{R}_{ij})
\eeq
The interaction between two clusters composing the giant molecule can be 
calculated as the double folding integral of ground state one-body densities
$\rho_{1(2)}(\bd{r})$ of heavy ions:
\beq
V(\bd{R}) = \int d{\bd r}_{1} \int d{\bd r}_{2}~
\rho_{1} ({\bd r}_{1}) \rho_{2} ({\bd r}_{2}) v({\bd s})
\label{unu}
\eeq
We employ the M3Y $NN$ effective interaction for the nuclear part of 
$v$ as described in \cite{smc98} to which we add a repulsive core
in order to take into account two major factors - the density dependence of 
the $NN$ interaction and the Pauli principle, which are important at distances 
corresponding to the overlap of the nuclear volumes. 
This choice is particularly useful for a molecular model in which the 
repulsive core prevent the re-absorption of the lighter fragment by the 
heavier one.

We consider that the nuclei composing the giant molecule 
are in their ground state with known quadrupole $\beta_2$, octupole $\beta_3$ 
and hexadecupole deformations $\beta_4$.

From what has been said above the interaction between the light cluster 3
and the heavy fragments remains unmodified when the bonding angle is
decreased. The reciprocal distances and orientations between the light 
cluster and the heavier fragments being freezed. 
Therefore its contribution to the total Hamiltonian
adds only a constant term. On the contrary, the interaction of the 
two heavy fragments 1 and 2 depends on the butterfly angle $\eps$ as can
be seen from the multipolar expansion of the potential
\beq
V_{12} = \sum_{\lambda_1,\lambda_2,\lambda_3,\mu}
\frac{4\pi}{\sqrt{(2\lam_1+1)(2\lam_2+1)}}
~V_{\lam_1\lam_2\lam_3}^{\mu -\mu 0}(R_{12})
Y_{\lam_1\mu}(\eps,0)Y_{\lam_2-\mu}\left (\frac{R_1+R_3}{R_2+R_3}\eps,0\right)
\label{2}
\eeq   
For small non-axial fluctuations(bendings), the potential in the neighborhood of 
the scission, or "molecular equilibrium" point $R_0\equiv R_1+R_2+2R_3$, 
gets a 
simplified form, provided we keep terms up to the second power in angle :
\beq
V_{12} = V(R_0) + {1\over 2}C_\eps\eps^2 
\eeq 
where the stiffness parameter reads
{\small
\beqa
C_\eps & =  & -{1\over 2}\sum_{\lam_1\lam_2\lam_3}
\left [\lam_1(\lam_1+1)\frac{R_2-R_1}{R_2+R_3}+
\left(\lam_2(\lam_2+1)\frac{R_1-R_2}{R_2+R_3}
+ \lam_3(\lam_3+1)\right ) \frac{R_1+R_3}{R_2+R_3}\right ]
V_{\lam_1\lam_2\lam_3}^{0~0~0}(R_0) 
\nonumber\\ 
&&-\sum_{\lam_1\lam_2\lam_3}\frac{R_1+R_3}{R_2+R_3}
\left (
R\frac{\partial V_{\lam_1\lam_2\lam_3}^{0~0~0}(R)}{\partial R}\right )_{R=R_0} 
\label{stiff}
\eeqa
}
Hence the quantized vibrational Hamiltonian of the giant tri-nuclear 
molecule acquires the form 
\beq
H_{vib} = -\frac{\hbar^2}{2{\cal
J}_{\eps}}\frac{\partial^2}{\partial\eps^2} + {1\over 2}C_{\eps}{\eps^2}
\eeq
The spectrum of the butterfly vibrations is given then simply by 
\beq
E_{\eps} = \left( n_{\eps}+{1\over 2}\right )\hbar\omega_{\eps}
\eeq
where $\omega_{\eps}=\sqrt{C_{\eps}\over J_{\eps}}$.

As one can see from Table I, in the tri-nuclear case the values of 
$\hbar\omega_{\eps}$ are only slightly smaller if we take into 
account Nuclear+Coulomb forces compared to considering only 
Coulomb forces. This was to be expected since in this case the 
distance between the two heavier fragments is already some fm 
beyond the top of the barrier and therefore the Coulomb forces 
are clearly dominating the interaction. This has to be contrasted to the 
case when we remove the lighter cluster, i.e. we insert $R_3=0$ and 
$A_3$=0 in (\ref{stiff}), and the two heavier fragments are brought
in touch. In this case the nuclear contribution to the interaction
significantly decreases the value of $\hbar\omega_{\eps}$.
Notice also the variation with the type of nuclear collective flow 
reflected in the type of inertia moment, the smallest values for 
$\hbar\omega_{\eps}$ being obtained for rigid rotations.

To have an ideea of the order of magnitude of tri-nuclear molecules 
life-times we evaluate the half-life of the corresponding di-nuclear 
configuration, when the lighter cluster is absent. 
Using the one-dimensional WKB formula 
\beq
\lambda = {\omega_R\over 2\pi}\exp\left \{
-2\int_{R_{t2}}^{R_{3t}}\sqrt{{2\mu_{12}\over\hbar^2}
\left ( D-{\hbar\omega_R\over 2} \right )}\right\} 
\eeq
for the decay rate of a metastable state of energy 
\beq
{\hbar\omega_R\over 2}={1\over 2}\hbar\sqrt{{1\over \mu_{12}}
\left ( \frac{\partial V_{12}(R)}{\partial R^2}
\right )_{R=R_{min}}} 
\eeq
in the potential pocket of depth $D=V_{12}(R_{max})-V_{12}(R_{min})$
of the two heavier nuclei, we obtain the half-life:
\beq
T_{1/2}=\frac{{\rm ln}2}{\lambda}
\label{halflife}
\eeq

The computed life-times for di-nuclear molecules  have large values($ \gg 
10^{-13}$s, according to Table I) and since 
the light particle has the tendency to delay the penetration  of the mutual barrier of
the heavier fragments, due to the attractive interaction with the heavier fragments, it is 
then justified to expect even larger values of $T_{1/2}$ for tri-nuclear 
molecules.

We choosed in this paper a linear configuration for the giant tri-nuclear
molecule. According to very recent calculations of penetrabilities in 
alpha ternary cold fission this configuration is reached at a certain 
step of the tunneling process \cite{smcg99}. It could be possible that 
before the beginning 
of mutual penetration of the multi-dimensional barrier, the giant tri-nuclear
molecule is found in a tri-angular quasi-equilibrium configuration.
In a forthcoming paper we will present the calculations of the molecular
collective spectra also for such a configuration but the general trends
should be same : considering an adiabatic scenario for the decay, the 
collective molecular modes will adjust slowly to the value of the elongation 
variable. As we mentioned earlier, the life-times of 
such nuclear molecules are suspected to be larger than 10$^{-13}$s, whereas
the life-times of the first molecular states computed in this paper are 
expected to be $\tau\approx{1 / \omega_{\eps}} >
10^{-22}$s. The gammas coming from the de-excitation of these states
should be observed before tunneling ended and therefore they should not
be pronouncedly Doppler-shifted. 

\section*{Acknowledgements}
One of the authors(\c S.M.) would like to acknowledge the financial 
support from DAAD-Germany.

\begin{table}
\caption{The quantum energy $\hbar\omega_\eps$ of the butterfly mode 
in the case of tri-nuclear ($^{10}$Be and $^{12}$C accompanied ternary 
fission) and di-nuclear configurations. $N+C$ signify 
calculations with nuclear and Coulomb forces, whereas $C$ only with 
Coulomb forces. Three types of the inertia moments are considered: 
a) experimental , b) Irrotational Fluid and c) Rigid Rotator. 
On the last column we listed the half-life  of the 
di-nuclear molecule.}

\begin{tabular}{c c c c c c}
 Splitting & \multicolumn{5}{c}{$^{96}$Sr+$^{10}$Be+$^{146}$Ba}\\
  & \multicolumn{2}{c} {Tri-nuclear} & \multicolumn{3}{c} { Di-nuclear} \\
Inertia moment  & $\hbar\omega_\eps^{N+C}$(KeV) & $\hbar\omega_\eps^{C}$(KeV)  & 
$\hbar\omega_\eps^{N+C}$(KeV)& $\hbar\omega_\eps^{C}$(KeV) & $T_{1/2}(s)$ \\ 
$J=J_{exp}$ & 1900.4 & 2040.4 & 1100.7 & 3377.3 & \\ 
$J=J_{IF}$& 1691.2 & 1815.9 & 880.9 & 2702.8& $\leq 5.5\times$10$^{-9}$\\ 
$J=J_{RR}$& 1014.4 & 1089.1 & 440.4 & 1351.4 & \\
\hline
 Splitting & \multicolumn{5}{c}{$^{96}$Sr+$^{12}$C+$^{144}$Xe}\\
 & \multicolumn{2}{c} {Tri-nuclear} & \multicolumn{3}{c} { Di-nuclear} \\
Inertia moment  & $\hbar\omega_\eps^{N+C}$(KeV) & $\hbar\omega_\eps^{C}$(KeV)  &
$\hbar\omega_\eps^{N+C}$(KeV)& $\hbar\omega_\eps^{C}$(KeV) & $T_{1/2}(s)$ \\ 
$J=J_{exp}$ & 1799.7 & 1917.3 & 791.4 & 3459.6 & \\ 
$J=J_{IF}$& 1611.5 & 1716.8 & 625.5 & 2734.5& $\leq 9.6\times$10$^{-9}$\\ 
$J=J_{RR}$& 996.0 & 1061.1& 312.75 & 1367.25 & \\

\end{tabular}
\label{table:1}
\end{table}

\end{document}